# ANALYSIS OF DESIGN PRINCIPLES AND REQUIREMENTS FOR PROCEDURAL RIGGING OF BIPEDS AND QUADRUPEDS CHARACTERS WITH CUSTOM MANIPULATORS FOR ANIMATION

Zeeshan Bhati, Asadullah Shah, Ahmad Waqas, Nadeem Mahmood

Khulliyah of Information and Communication Technology

International Islamic University Malaysia

## ABSTRACT

*Character rigging is a process of endowing a character with a set of custom manipulators and controls making it easy to animate by the animators. These controls consist of simple joints, handles, or even separate character selection windows.This research paper present an automated rigging system for quadruped characters with custom controls and manipulators for animation.The full character rigging mechanism is procedurally driven based on various principles and requirements used by the riggers and animators. The automation is achieved initially by creating widgets according to the character type. These widgets then can be customized by the rigger according to the character shape, height and proportion. Then joint locations for each body parts are calculated and widgets are replaced programmatically.Finally a complete and fully operational procedurally generated character control rig is created and attached with the underlying skeletal joints. The functionality and feasibility of the rig was analyzed from various source of actual character motion and a requirements criterion was met. The final rigged character provides an efficient and easy to manipulate control rig with no lagging and at high frame rate.*

## 1.Introduction

The process of animation a virtual character is long and tedious work. There exists huge number of rigs, tools, software's which are very advance and functionally provide a good standard rigs with ability to do tons of things. These software although are very efficient and advance but they don't always satisfy the needs of computer animator and so usually a custom process of endowing an object with a set of controls is done to achieve greater control over the animateable character. This process is normally termed as *Rigging*. Generally defining, Rigging is a fundamental part of the animation, where various custom controllers are attached to each skeletal body part. These controllers and manipulators usually consist of simple joints, locators, selection handles, spline curves, or even an independent graphical user interface (GUI) for control selection [1]. By connecting a rig to a model in a process called binding, the model mimics the motions of the rig like a puppet. The boredom of manually doing this process for each character and object in a project makes the pipeline of character animation more time-consuming, difficult and problematic[2].

A good character rig is created according to the needs and principle requirements of the characters motions. A biped rig having controls that make sense, be easy to understand with controls placed in accurate location and work in a consistent manner, will immensely help and aid the animator to bring the 3D virtual character to life with easiness and believability[3]. Often it has been seen that even the animator, riggers and technical directors also tend to forget about essential and important little things that makes animating with a rig easier and streamlined process.

This research work implements a template based skeleton generation mechanism called widgets, for Biped and Quadruped character types, based on the actual anatomy of each character type. The foundation of scripted rig building is definition of the number and location of joints. Then this skeleton is automatically rigged according to the various standards and criteria researched and discussed with custom controls and manipulators. This system of automation will provide a practical solution to the real life problem of character rigging and animation.

The proposed system of automated rigging of Biped and Quadruped characters with custom manipulators, is achieved by procedurally generating the entire system with very minimal user intervention. The biped and quadruped character rigsare automatically generated with all custom selection controls basedon inverse kinematics (IK) and forward kinematics (FK). This automated technique of procedurally generating entire character's Rig makes the process of rigging and character type a stress free and timesaving. This system will facilitate the novice character rigger and animator greatly by aiding to create and use an advance character rig through very few user intervention.The major benefits of using procedural technique to automatically create biped or quadruped rig is that the time spent to build a system for dynamic motion control and deformation system is decreased  by scripting the entire process and generating the rig through a very few mouse click.

This widget based system provides a practical solution to the real life problem of character rigging and animation.This work is an extension of previously presented work on Biped Rigging [4] and for Quadruped Rigging in [5].

## 2.  Related Work

Auto Rigging and Skeleton generation: Most of the work on automated rigging focused on various techniques of extracting the skeleton from a given mesh. Repulsive force fields were used by Liu et al. [6] to find a skeleton.Whereas, Katz and Tal [7] suggested extraction of skeleton as an application through surface partitioning algorithm. The technique used by Wade [8]is to approximate the medial surface by finding discontinuities in the distance field, but they use it to construct a skeleton tree. The proposed algorithm by Pantuwong[9]uses high-curvature boundary voxels to search for a set of critical points and skeleton branches near high-curvature areas.Whereas, in a different approach, Pantuwong proposes a technique of automatically generating inverse kinematics based skeleton using skeleton extraction from the volume of character mesh [10]. In contrast, Baran develops a prototype system called Pinocchio where he implements a method of generating Skelton and automatically attaching it to the character's skin/mesh [11].

Another common technique is template fitting and matching techniques for skeletal generation. This approach provides accurate skeletal generation and matching to the original mesh[12]. Majority of the work using this technique focusses on human characters for segmenting the mesh according to the human anatomy [13]. Anderson [14]fit voxel-based volumetric templates to the data. On the contrary, Liu and Davis discuss a new facial rigging system that hybridizes several of the traditional rig interfaces [15]. Whereas, other several other researchers have worked on creating an automated rigging system targeted specifically on face rigging include [16], [17], [18], [19] and [20].

## 3.  Basic Principles of a  rig

To create an advance production standard animation rig, it is very vital to understand the actual requirements and the types of motion the character is going to perform. The auto rigging system developed in this paper concentrates on the following overall rig criteria.These are the few fundamental norms that have been followed in this system but are not limited only to these.

### 3.1 General Rig criteria:

*i.* The rig should be consistent; meaning that when animating the controls should not break the rig apart or follow transformation in an unorthodox manner.

*ii.* The rig should have Predictable behavior and all the controls should behave and operate exactly the way they are intended to work.

*iii.* The control structure should be as simple as possible and not cluttered with multiple controllers and manipulators hanging about for the animator.

*iv.* The Rig should be easy to use with minimum number of controls and maximum functional management.

*v.* The rig has to be lightweight and fast in interaction.

### 3.2 Animation criteria:

i. The rig must be built while considering that how the character should act and perform, as to bring out his personality.

ii. It is very essential to know what the director want from the character and what the storyboard is. What are his requirements as to the motion types the virtual character is performing, i.e. jump, fall backwards, martial arts fighting, swimming, flying, etc. All these require special consideration while rigging with special setup.

iii. It's also important to get feedback from the animator regarding his needs and requirements of the controls and functionalities of the rig. After all it will be the animator who will eventually use the rig

## 4. Guidelines for developing A rig

On the basis of the above norms, a set of guideline have been proposed for the creation of a functionally advance bipedal and quadruped rig. It is to understand that a functionally great animation rig is determined by the ability, freedom and range of all possible movements that are achieved using it with least amount of effort. Hence having tons of controls to manipulate various body parts does not yield a rig to be of highest rank. Therefore, the best way to develop a functionally valuable rig is by logically and artistically thinking about all the movements and actions a character performs in real life and then building a rig so that it is able to mirror those gaits and motions with minimum efforts and control manipulation. Hence, the following design guidelines are proposed and were developed through monitoring and analyzing the real life movements of a human character as shown in Figure 1.

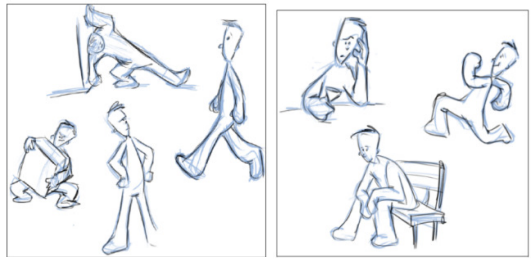

Figure 1: Human child in their natural poses [21].

For this research work, the motion reference of 2 Children is used, as it was analyzed that children perform wide range of bizarre motions and extreme gait poses specially when they are playing and having fun. Whereas the motion of a grown adult is always predictable and driven intentionally so to understand the pure flexibility of human body the best reference would be a child in play time.

1. Set the rotation order of all the controllers and manipulators in a way that makes sense with properly aligned XYZ axis. A proper setup of rotation order helps to avoid and prevent the **Gimbal Lock.**

2. Each control in skeletal hierarchy should be able to use Maya's pick-walk feature that allows the animator to select the controls - that are in hierarchal order, by using simple up and down arrow keys.

3. The rigger should avoid creating two controls on a same body joint, performing almost the identical transformations resulting in two unnecessarily matching motion curves. For example, two controllers at wrist joint controlling the hand rotating using FK and IK, this both rotating hand thus generating two curves individually. Similarly, controllers found around the main pelvic area can be up to 3 distinct manipulators for hip, root and upper torso rotation. All these controllers affect the upper torso and hip area doing almost exactly the same thing. Therefore,there should be no redundant controls.

4. The curve based manipulators and controllers should be visually unique and identifiable by their shape and color. If two controllers are of exactly same shape & color i.e circular blue, then it really complicates the animator every time, regarding the purpose of each of them. For example, translation based controllers can be of arrow shape, whereas rotation can have circular shape with color segregation on right and left side.

5. All the rotation and translation values should technically be accurate and follow a natural direction of motion. For example, the controller should give a positive rotation values when rotated forward and similarly a negative values should be given when controller moved in opposite or backward direction.

6. Another common mistakes made by riggers is setting the limits on custom attributes. The custom parameters such as Foot-Lift or Finger-Curl should never have maximum or minimum limits from -1 to 1, or 0 to 10. The rig should ensure that the custom parameter should have same familiar motion curve in the Graph Editor for all the attributes and transformation, therefore a custom limit of -180 to 180 is more appropriate.

7. Rigs needs to be fast, and effective, therefore it's always recommended to use 'Nodes' instead of 'Expressions', as nodes are more faster in calculation as compared to expressions. For example, to calculate the distance between two points, use distance nodes instead of writing an expression. Try thinking outside the box, i.e., a rendering or invert node might be able to solve the basic calculation that isprerequisite in a rig or a RGB - XYZ blend color node can be used instead of blending between two constraints.

## 4.1 The criteria  followed in this System

Each procedural rig automatically generated by the system, is based on complex and advance set of controllers and manipulators compiled together in a user efficient and with easy to uses functionality. The rig is created while ensuring all the possible body movements and requirements of an animator from an industry as discussed in this paper.The basic system pipeline fallows the following criteria:

- Creates joint hierarchy according to the human and quadruped anatomy.
- Rigs all parts of the character automatically and cleans the scene for faster playback.
- All body parts are rigged separately and independent of each other and grouped under separate nodes, so that each body part i.e. left arm, right arm, spine, neck, etc. can be easily taken apart and deleted or detached from the main rig without effecting the entire rig. This gives the isolation functionality for example, having just one arm or one leg in the character rig.
- Rig is created according to the various requirements of the human and animal locomotion types with various range of gaits.
- All principle of a standard character rig are incorporated in the rig including increasing the length of body parts, having the ability to stretch and squash the rig.

## 5.The overview of the System

The basic architecture of the proposed system is initially based on creating a template based widgets for each biped and quadruped character types. Thenthe user is given control to adjust the basic widget structure to fit in the 3D modelled character type. Afterwards, joint based skeleton is automatically extracted and constructed on underlying widget locations, and finally generating the complete rig automaticallyon the skeletal structure.It is to note here that the widget structure of biped and quadrupeds is completely different as shown in Figure 2.

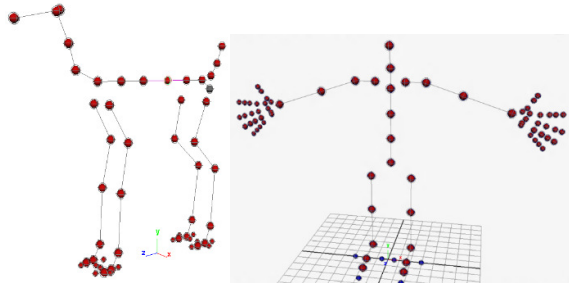

Figure 2: Widget structure for a quadruped character

The widgets are basically NURBS spheres with circular curves, which then are hierarchally parented to each other forming a biped or quadruped structure based on predefined location coordinates. The basic architecture of the proposed scheme is shown in
Figure 3.

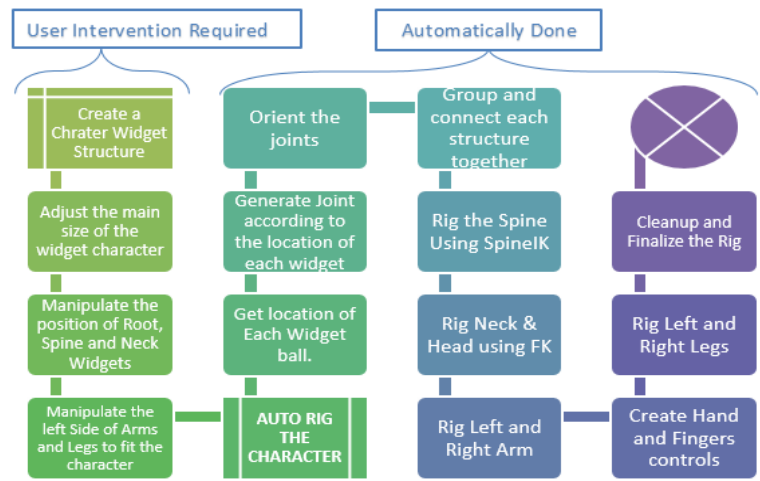

Figure 3: The Process flow of the auto rigging system

There are two stages of the pipeline. In first part the widgets are created automatically procedurally on a predefined & calculated location. The user adjusts the position of each widget unit, which represents a single joint. In the second stage the entire character rig is created procedurally by the system through a single click of a button.

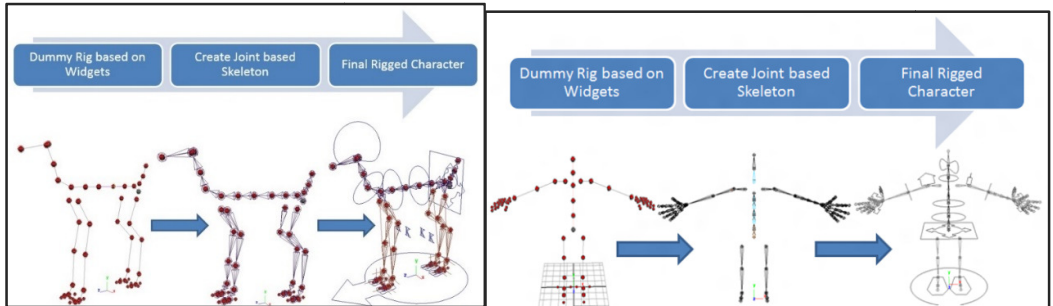

Figure 4: The basic Pipeline of the auto Rigging System for Biped (left) and Quadruped (right) character types.

# 6. The Widget System

The process of rigging a characters starts with creation of quadruped widget structure through a GUI system shown in Figure 4. Through theGUI, the rigger has the control to create the widget system of each body part independently or simply for an entire character. Creating an independent individual body part widget is used to create rigs for unorthodox or nonhuman like character. This widget is placed according to the hierarchy of a quadruped skeletal structure instead of joints or bones. The user simply adjusts the widgets according to the size and shape of its quadruped character[4]. This is the only user interaction part needed in the rigging process.

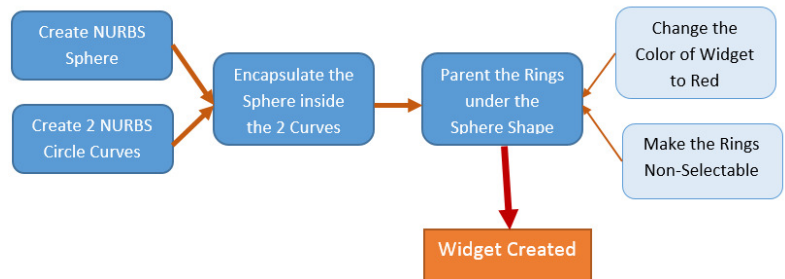

Figure 5: Widget Creation Process

## 6.1 Widget creation proccess

A basic widget unit is created by simply using a NURBS sphere object. This sphere is then encapsulated with two circular rings of curved lines. These curved rings are placed over the sphere object and parented under it, to form a single selection point as shown in Figure 6. Their pivot points are also centred with respect to each other. Each of this widget unit represents a single joint location in a skeletal hierarchy as given in Figure 8. Then, each widget unit is procedurally connected with another relevant unit in its hierarchy, using a spline straight line.

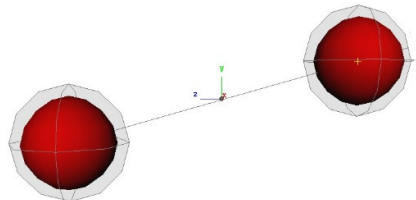

Figure 6: A widget unit connected with another widget unit through a spline curve.

Initially, the pivot location of each widget unit is determined. Then based on these coordinates, a line spline based straight line is drawn with linear degree, having only two vertexes. Each Control Vertex (CV) of the line is located at the centre of corresponding widget unit. Then, each control vertex is selected and a soft-body cluster is created on that vertex for soft deformation of the spline curve. The cluster handle is then parented under the relevant widget unit. This unique technique allows the widget unit to be selected and translated with the line following and automatically deforming appropriately to follow and match the location of widget unit. This effect creates a perfect scenario for a Bone-Joint relation. This entire process is illustrated in Figure 7.

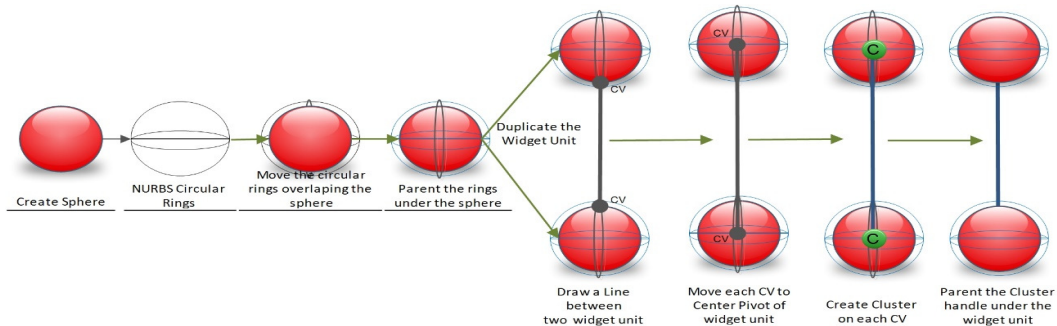

Figure 7: Process illustrating the widget unit creation

Using this process, the entire widget hierarchical structure is created. The exact location of each widget unit is per-calculated manually, according to the quadruped skeleton. Each widget unit is duplicated and moved at these joint locations, procedurally. Then, each joint is connected with spline based linear degree curved lines, as discussed previously. Finally the right side joints, for example, the right front leg widgets, are constrained and mirrored, to follow the exact reverse transformation value from its left side widgets. This allows the user to only manipulate and modify the left side of the widgets and the right side widgets will automatically adjust themselves, creating an exact mirror effect. This unique technique greatly reduces the user's effort and saves time, as only few joints need adjustment, according to their custom quadrupal character size and proportion. A simple widget based hierarchal joint chain for two hind feet's are shown in Figure 8, they have the same auto-adjust functionally on both sides.

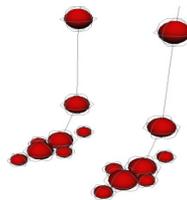

Figure 8: Widgets for Left Hand and Feet's with pivot control

## 6.2 Skeletonization Process

The joint based skeletal structure is generated through finding transform location of each widget sphere. Based on the widget sphere location, the joint are created and moved to that exact location procedurally. Hence the underlying widget structure is quite important and controls the placement and hierarchical order of the skeleton.

## 7. SpineRigging

The development of the system starts from torso or spine. First the reference images were analyzed to determine the range of movements and possible solutions. The spine determines and illustrates the entire body pose,as it can see from the Figure 9. It is the most crucial part as it holds all the different body parts together and thus becomes the origin of their motion.

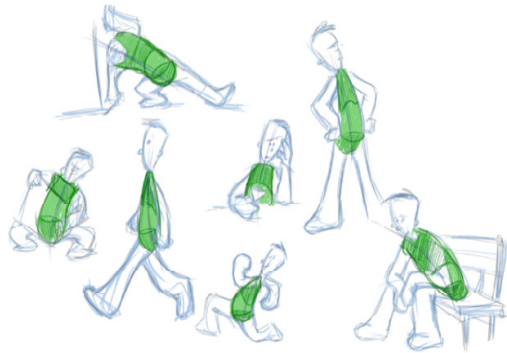

Figure 9: Torso motion references.

## 7.1 Root Rigging

The root controller manipulates the entire upper body portion of a character. The root controller must have all its transform values to zero, at a default pose. The freeze transformation feature of Maya does not work on root joint, as it is the top most joint in a hierarchy. Therefore, one way to achieve this it to Group the root joint to itself. Then move the pivot point of the group to center of root joint and parenting it under the root controller.

## 7.2 Spine Movement Objectives:

After analyzing the reference images and videos, following set of objectives have been concluded for spine rig

1. The controls should have the rotation of hips and shoulders
2. The controls should allow the rotation in all axis – Bend, Side to Side ,and Twist
3. The controls should provide independent motion of shoulders and hips.
4. The spine rig should have the functionality for relocation of pivot.

The algorithm 1 given is the pseudo code for creating spine rig. This process is illustrated in Figure 10.

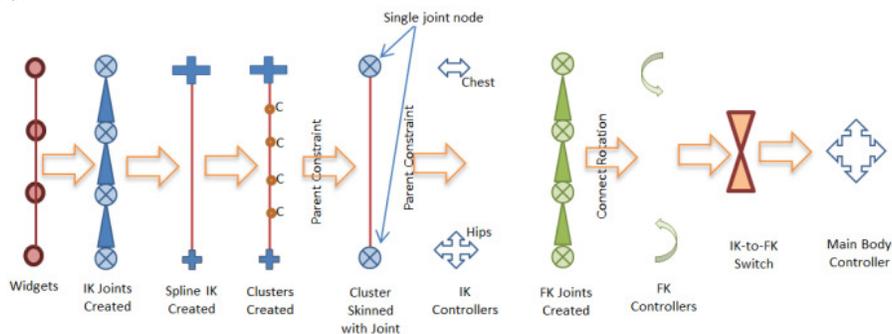

Figure 10: Process of creating Spine rig

Algorithm 1: Pseudo Code for the process of creating the spine rig

1. READ position of hip widget and all spine widgets
2. CREATE hip and spine joints at their corresponding position of widgets
3. RENAME all the joints
4. CREATE Spline IK solver from spine-1 to last spine joint
5. ATTACH skinCluster between hip_joint, last_spine_joint and IK_curve

6.   CREATE curve based controllers at Hips and Chest
7.   CONSTRAINT Hip controller to hip joint
8.   CONSTRAINT Chest controller to last spine joint.
9.   CREATE FK joint chain based on widgets location
10.   CREATE Curve based controllers for FK rotations
11.   CONNECT the FK joints rotation to FK_controllers
12.   CREATE main body controller
13.   PARENT all the controller under it body_controller.

## 7.3  Stretchy Spine

The ability to stretch a joint chain is extremely useful in animation. To make a joint chain, the distance between the joints is determined and how far they are from their locator. The spline curve used by spine IK is used to determine the final value of joint scale. The distance between two joints shown in Figure 11in the spine rig is determined by measuring the arc length of the curve at current position $C_l$ divided by its original rest pose length $C_o$. The equation used to calculate the scale factor $S_f$ is

$$S_f = \frac{C_l}{C_o}$$

Each joints scale is then set to Sf using procedural expressions.

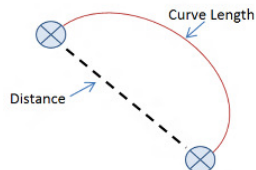

Figure 11: Process of creating spine rig.

Figure 12shows the final completed rig of biped and in Figure 13 Quadruped spine rig is shown. Each type of rig contains custom manipulators and controllers for motion control and deformation of the spine region.

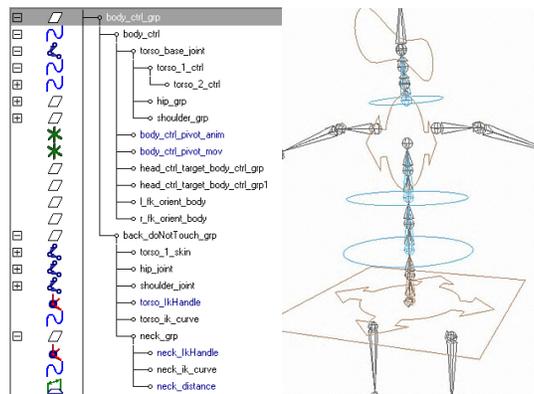

Figure 12: (a): Hierarchical node structure of the torso rig. (b) Final Completed spine Rig

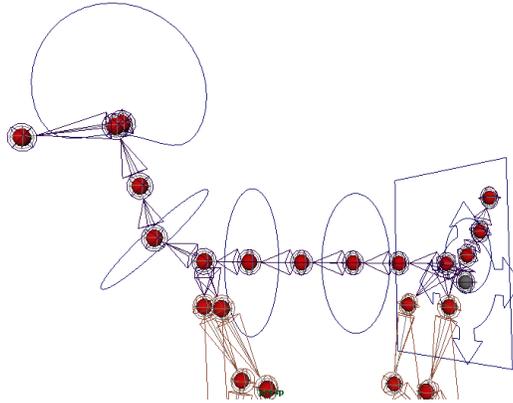

Figure 13: Final completed Spine rig of quadruped

# 8.HEAD and NECK

Head and neck are the key body parts in rigging as their relationship with each other expresses the attitude of the character. Figure 5 shows the final rig controls for the head and neck region. The neck; having multiple joints, is controlled through inverse kinematics (IK) based on spline curve. This gives us the smooth curvy bend around the neck joints. Contrary to neck, the head is controlled using simple forward kinematics system. The pseudo code of head and neck rig system is given in Algorithm 2.

Head and neck are the key body parts in rigging as their relationship with each other expresses the attitude of the character. Looking at the references images following requirements have been set for the head rig.

1. Head rig needs to be able to orbit side-to-side and look up and down.
2. Head rig has to lean and move side-to-side also.
3. Head rig needs to be able to move forward and back
4. The rig should have the feature to compress and extend
5. The movement of head should have the control to be independent of shoulder and body movement.

Algorithm 2: Pseudo Code for the process of creating the Head and Neck rig

1. READ position of Neck and Head widgets
2. CREATE Neck and Head joints at their corresponding position of widgets
3. RENAME all the joints
4. CREATE curve based controllers for Neck and Head
5. CREATE Spline IK solver from neck_base to neck_end joints
6. CONNECT twist attribute of IK to neck.twist attribute.
7. CONNECT the rotation of head controller to head joitn
8. PARENT head controller to neck controller
9. CONSTRAINT Neck controller to Chest Controller
10. ADD and connect attribute to switch Neck-Chest constraint ON or OFF.

# 9. Arms Rigging

## 9.1 Requirements for arms

Going through various references to analyze natural moves that an arm is able to express in a natural behavior, following key requirements are summed for creating an arm rig.

1. For free form waving and gesturing of arm, Forward Kinematics (FK) setup is required.
2. Inverse Kinematic setup for placing hands on the table or in the ground, or holding on to something, or while sliding the hand along a trajectory.
3. Providing an Elbow Locking mechanism for the ability to place elbows on table.
4. Shoulder control to facilitate biomechanically correct arm movement.
5. The rotation of the arm should have the ability to be independent from the shoulder and the body.
6. The arm rig should have the ability to stretch.

## 9.2 Arm Rigging Technique

The inverse kinematics (IK) system used in the arm, automatically calculates the angle of an elbow based on the distance between the wrist and the shoulder as shown in Figure 14.

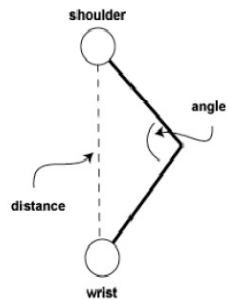

Figure 14: The distance & angle of biped arm joints used in IK setup

Algorithm 3: Pseudo Code for the process of creating the Arm rig

1. READ position of Arm widgets
2. CREATE Arm joints at their corresponding position of widgets for FK motion
3. RENAME all the joints
4. CREATE locator at the shoulder joint of arm
5. PARENT the Shoulder_joint to the Shoulder_locator
6. PARENT the Shoulder_locator to the lastSpin_joint
7. CREATE another 2 locators at Shoulder_joint and lastSpine_joint
8. RENAME them to Spine_orient and body_orient respectively
9. ORIENT Constraint the Spine_Orient and body_orient locators to Shoulder_Locator
10. ADD Attribute to control the switch between the two orients.
11. Create Curve based controllers and connect them with arm joints for FK rotation
12. CREATE Arm joints at their corresponding position of widgets for IK motion
13. CREATE ikRPSolver IK handle between ikShoulder_joint and ikWristJoint
14. CREATE Curve based controlers at wrist and near elbow
15. Connect the ikWrist_ctrl to ikHandle and ikPoleVector to ikElbow_ctlr

### 9.3  Streatchy Arm Setup

Creating stretchy arm setup is not as simple as scaling the joints because this will create an abnormal behavior specially when bending the elbow. In order to solve this issue, the proposed technique is to first find the actual distance of the arm from shoulder to wrist when the arm is at full length stretch as shown inFigure 15. Then when the distance of the controller is increased and is greater than the distance of upArm and lowArm joints then, the length of the joints is increased to create the stretchy effect.

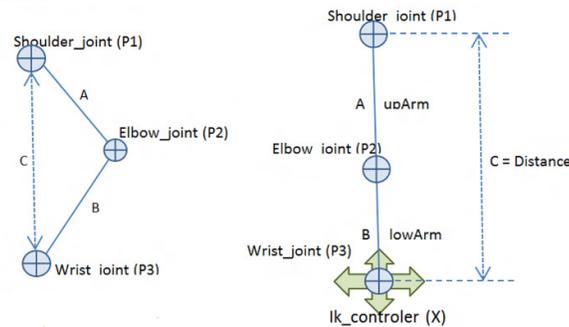

Figure 15: Default position with the scale not taking affect, because distance c is less than a + b.
Start scaling the joints using ikcontroler (x), now that distance c is equal to or greater than a +b.

Algorithm 4: Pseudo Code for the process of creating the stretchy Arm rig

1.  A (Length of UpArm) = Distance between P1 to P2.
2.  B (Length of loeArm) = Distance between P2 to P3
3.  C ( Full Length of Arm) = A+ B
4.  X = Controls the Streatch Factor of the Arm from P1 to P3
5.  if X < C then
        UpArm.Scale =1
        lowArm.scale =1
6.  else if X > C then
    a.  upArm.scale = x
    b.  lowArm.scale = x
7.  End If

### 9.4  Elbow Locking

The ability to lock characters elbow in certain situations is extremely necessary in animation. Since a working mechanism to stretch the arm has been developed in previous section, by measuring the distance between the shoulder and wrist joints of the arm, and when the joint chain reaches to its maximum length then start scaling the joints. Using the same methodology for elbow locking but instead this time the joints stick or stretch towards the elbow controller. The simple implementation logic is to measure the distance between the joints and the elbow, and then tell the joints to scale according to that new distance value as shown in Figure 16.

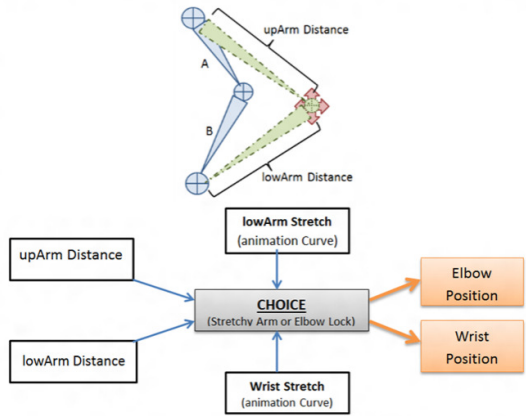

Figure 16: Elbow Locking and the node structure for the choice function

Then the animator will simply be given a choice to either stretch the arm from wrist or using the elbow.

## 9.5 Twistable Elbow

The twisting is the most essential effect of natural behaviors that has to implement in a rig, as this is the motion that happens almost naturally, and most of the time person is unaware of it. For example when the wrist is rotated sideways, actually it's the forearm that twists from the elbow to wrist and causes the sideways rotation of the hand. To simulate the twisting of arm joints in this rig, a sub-joint chain system is created between the elbow joint and wrist joint and spline-IK system is used to create the twist function much similar to that of spine rig system as illustrated in figure 8.

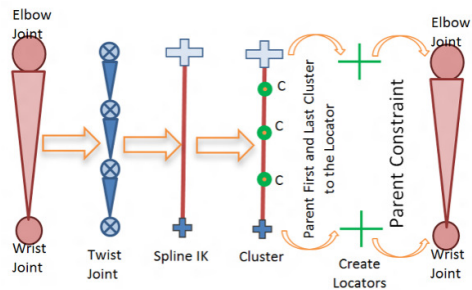

Figure 17: Twistable Elbow implementation

This system has an independent and isolated functionality from the rest of the body and so the entire arm rig can easily be deleted or modified without affecting the rest of characters rig in any way.

The Figure 18shows the entire hierarchy of the arms rig with various controls and constrains set up in a independent hierarchal system. This system has an independent and isolated functionality from the rest of the body and so the entire arm rig can easily be deleted or modified without affecting the rest of characters rig in any way. The final version of bipedal arm rig is shown in Figure 19, with forward kinematics and inverse kinematics based setups.

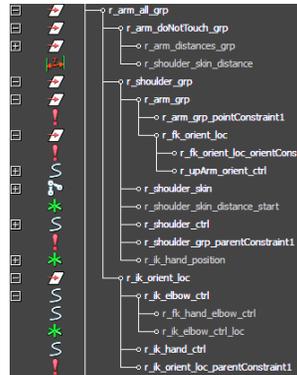

Figure 18: Hierarchal structure of the arm rig.

## 9.6  Fingers Rigging

Fingers are the most overlooked part of a biped character's motion, when animated. The fingers convey just as much emotion and intensity as the face expressions do. Hand gestures are often regarded as the punctuation of a character's body language. That's why the hands need careful thought and consideration.

[http://bryoncaldwell.blogspot.com/2008/04/hand-poses-galore.html]

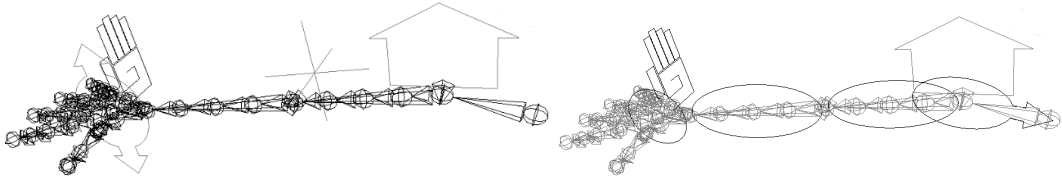

Figure 19: The IK (above) and FK (below) bipedal arm rig setup

Every animator will want to vary the shape of the fingers a bit. They will do it for various reasons: the character is doing something specific, the hands are moving quickly and they want to create a "smear" shape, the character is pressing down on something, etc. There are an infinite number of reasons as to why an animator would want to animate individual joints, and an infinite number of hand poses they should be able to create.

Let's create a list of the controls:
• Curl
• Thumb curl
• Scrunch
• Thumb Scrunch
• Relax
• Cup
• Spread
• Mid-Spread
• Thumb-spread
• Twist
• Lean

For the curl parameter of the finger, set driven key technique is used to set the manual nonlinear key on the custom curl parameter.  The curl attribute is set to 0 with all finger joints at default value with fingers in a straight orientation. Then a key is set, later rotationof all the joints of each finger of hand to create a curl, as shown in Figure 20, and a key is again set at 100 for the curl attribute. Now using this custom curl parameter, the animator can easily manipulate and control the curl feature of fingers.

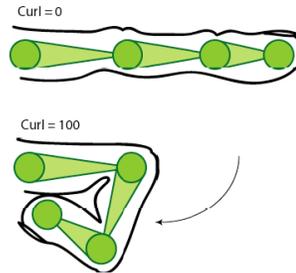

Figure 20: Finger Curl with 0 being default position and curl is set to 100 for full finger curl.

The Crunch parameter allows the finger to bend in an opposite direction as opposed to curl, where the fingers are bent in forward natural rotation. The crunch occurs when fingers or hand presses down on a table as shown in Figure 21. There are two ways of achieving this crunch effect, first is to create an IK handle on each finger from its base to last joint, and use that IK to achieve this effect. The second technique used in this system, is by grouping each fingers joints and parenting it to previous joint in hierarchy. Then moving the pivot of each group it the centre of finger joint, and using that group node along with set driven key approach to create the crunch effect. This group approach allows us to create number of other finger control parameters, with mixing few other techniques.

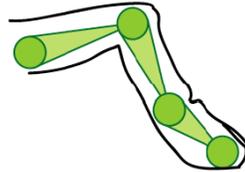

Figure 21: finger Crunch behaviour, occurs when hand is pressing down on a table.

## 10. LEGS RIGGING

Finally the Legs of a character are rigged. The legs primary responsibility is to actually provide the forward or reverse motion, caring the body with it. Nevertheless, as a matter of fact they propel more than just locomotion; legs gaits convey the essence of force, pressure and the structure of entire body movement. Following are the summarized requirements for the leg:

1. Almost 99% of time the character feet will need to be planted on the ground and the feet will drive the motion of entire leg.
Therefore Inverse Kinematics system will be used.
2. At certain unforeseen times the character needs to let the legs flow freely of example when falling, rolling over on a chair,swinging, and etc. so forward kinematics is also implemented.
3. To get that feeling of weight and pressure on character a footpivot and foot rolling system is required.

The rigging system for the leg is actually quite simple. The inverse kinematics (IK) system used to automatically calculates the angle of knee joints based on the distance between the foot and the upleg joints.The basic leg architecture shown in Figure 22involves IK and FK leg setup and the user gets to choose which system they will use. The pseudo code of procedurally creating leg rig controls is given in algorithm 3.

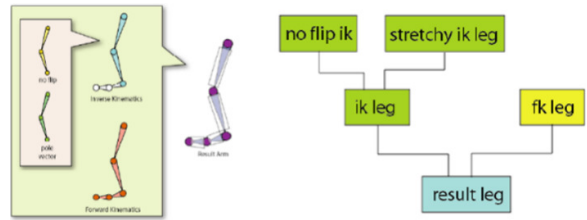

Figure 22: IK & FK Leg setup

---

Algorithm 3: Pseudo Code for the process of creating the Leg rig

---

1. CREATE IK_Handle between UpLeg_joint and Ankel_joint
2. CREATE IK_Handle between Ankle_joint and Ball_joint
3. CREATE IK_Handle between Ball_joint and Toe_joint
4. GROUP all three IK_Handles together
5. CREATE a Curve based leg controller
6. PARENT the Group under the leg controller
7. GROUP Ankel_IK_Handle to itself
8. MOVE the pivot of the group to Ball_joint
9. GROUP Ball_IK_Handle and Toe_IK_Handle together
10. MOVE the pivot of group to Ankel_joint
11. GROUP the Toe_IK_Handle to itself
12. MOVE the pivot of the group to Ball_joint

---

The algorithm 3 illustrates the basic steps in creating the IK based leg for the quadruped character. The grouping system allows for easy foot roll, ankle roll, toe lift and ball lift functionality. The rotate attribute of the group can then be connected to the custom attributes added to the Leg_controller for easy selection and manipulation of the foot. The stretchy leg effect is created using the same algorithm 4discussed previously. Figure 23shows the final rigged leg controls of a quadruped character.

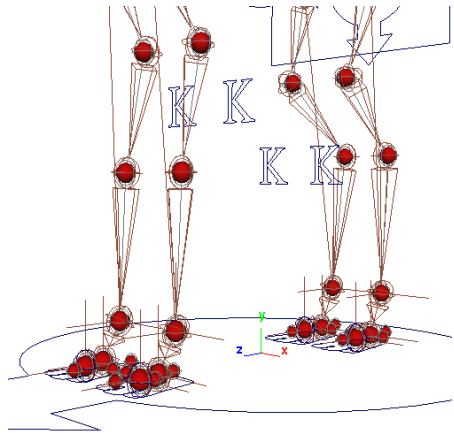

Figure 23: The IK and FK based leg rig controls of quadruped

## 11.Results

The auto rigging system for quadruped character has been tested by creating multiple rigs for various types of characters.Thefunctionality and the dependability of the rig are extremely efficient as compared to other freely available rigs on the internet.Finally the algorithms and procedures were compiled to create a full working plugin system for MAYA software. Thefinal rig created by the systemis tested on multiple quadruped and biped character types. The system proved to be extremely flexible and expandable to multiple characters of different size and proportions.

As a result the following key features have been achieved for the auto rigging system of quadruped characters.

- The rig can be created in any standard pose.
- The right side of the widgets is automatically mirrored reflecting the position of the Left side of the character.
- The entire body rig is independent and isolated from other parts.
- All the body parts are rigged automatically according to the animator requirements.
- Seamless matching from FK & IK switching is performed using the technique discussed in [22].
- As the rig has been designed in a structured manner thus it provides the functionality of mirroring the characters poses and also saving the poses and transferring the poses from one character to another as the underlying architecture is the same.
- The Leg and Arm rigs have the ability to stretch along with the ability to lock the Knee or Elbow movement.
- Extremely fast and clean rigs, with minimum no of nodes and expression for real-time feedback in viewport.

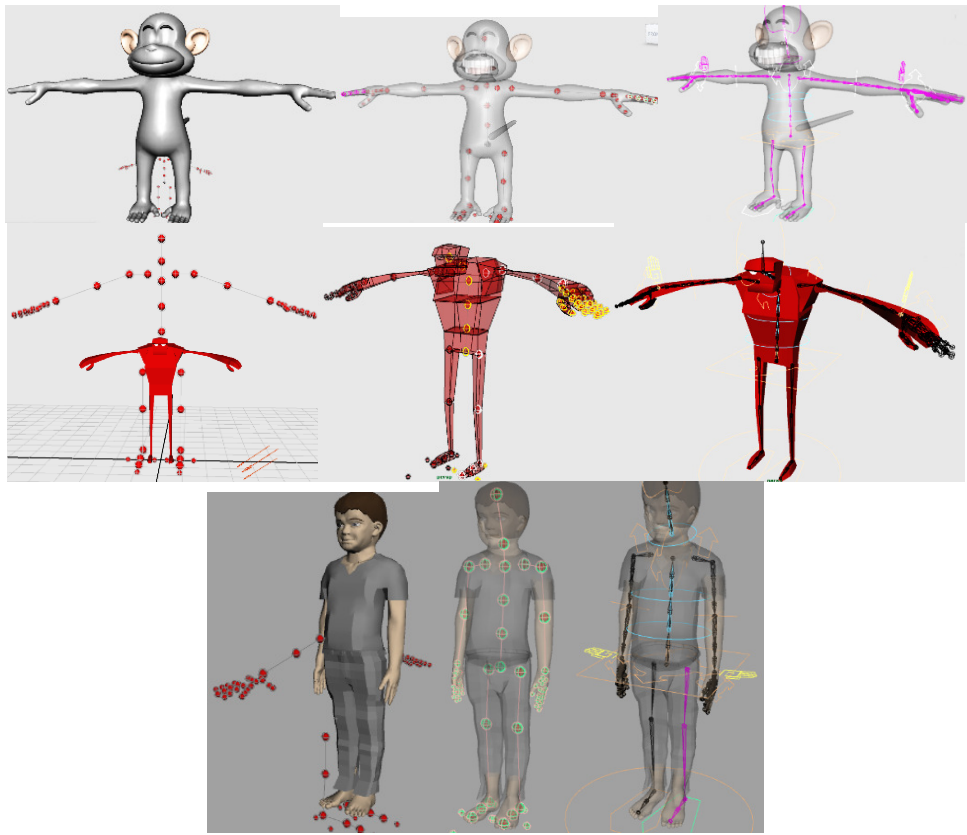

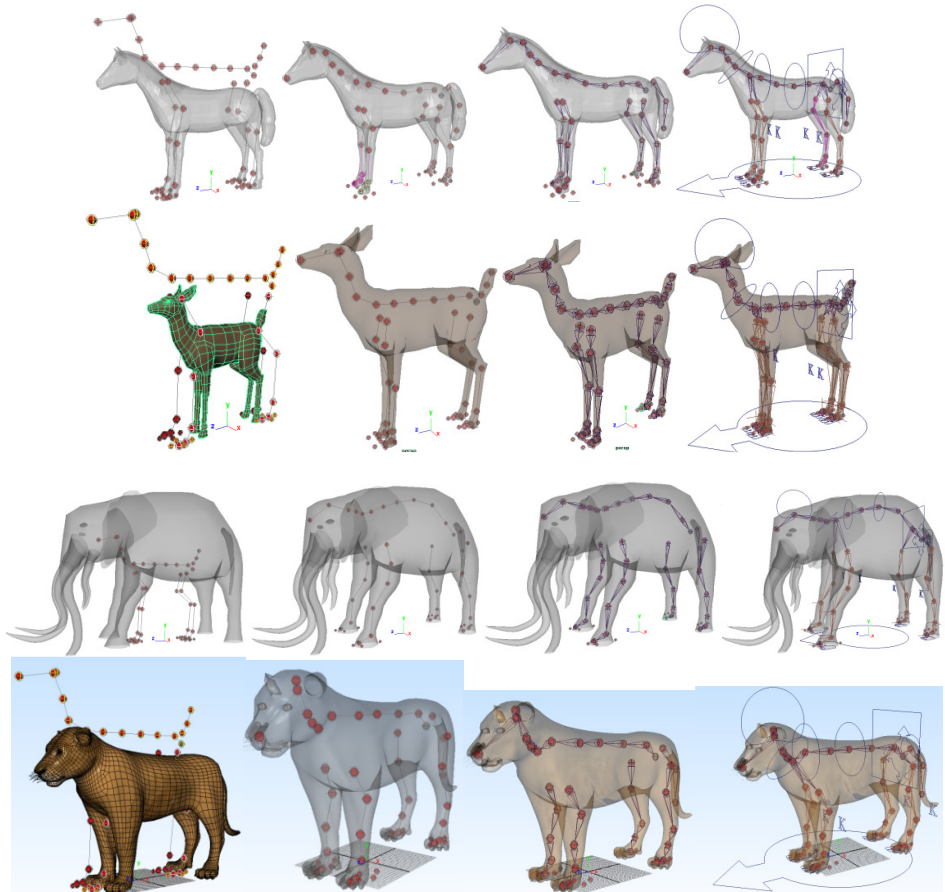

Figure 24: Final results of procedurally rigged biped and quadruped characters

## 12.Conclusion and Future Word

This paper discusses a new technique of generating a template based skeleton using widgets and then creating a fully functional automated quadruped rig with manipulators according to the basic principles and requirements of a standard quadruped rig. There are lots of resources available on internet regarding the quadruped rig yet none of them are concise and meet the need of an animator, moreover, none of the standard rigging requirements has been reported so far in literature on the same. In this paper a firmpolicy for the character rigger and animator has been provided through deep analysis of quadrupedand biped motion.The rules and principles of creating any type of character rig was obtained formextracting all the possible gait types from various video sources. Then a list, highlighting the key motion types and requirements has been discussedto aid the rigger and also so that they can be used as a reference guide. The working algorithm has also been discussed to implement the various rig types along with detail illustrations of the rigging process. The system is tested by implementing the automated rigging system on various biped and quadruped character types. These results show that, this template based widget system is very flexible and can be easily fitted on different 3D virtual characters. Once the widget structure is fitted according to the characters body size and proportion, then the user simple clicks on a button to generate joints based on widget location and finally generating a fully procedural rig. Finally a GUI is presented that would enable the animator to easily select and manipulate the character.

# Acknowledgment

This work is partially funded by Ministry of Higher Education Malaysia (MOHE) under Commonwealth Scholarship and Fellowship Program, Ref: KPT.B.600-6/3 (BP3173731) 2012-2014.

# 13.References

# Appendix

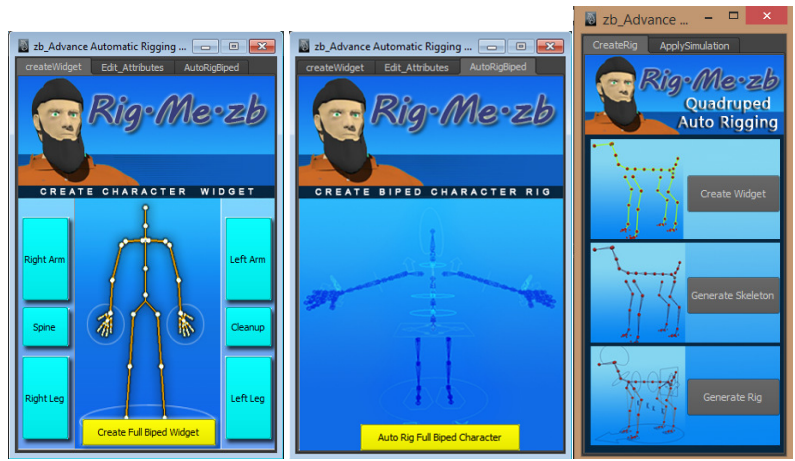

Figure A: GUI for creating auto rigging system fir biped and Quadruped character types

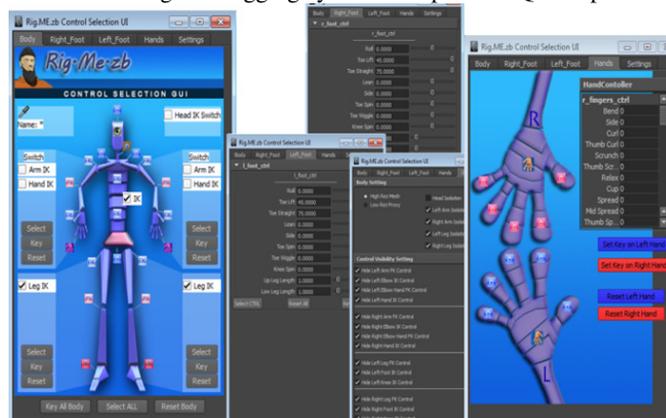

Figure B: GUI for controlling and manipulating the auto generated biped rig

# Authors

**Zeeshan Bhatti**

Mr. Zeeshan Bhatti is a PhD(IT) researcher in the field of Computer Animation at Kulliyyah of Information and Communication Technology, International Islamic University Malaysia (IIUM). His current area of research is in the field of Computer Graphics, 3D Animation and Modelling, procedural animation and simulation techniques, Motion Analysis with Gait categorization, and Multimedia Technology. His PhD research topic is "Oscillator Driven Central Pattern Generator (CPG) System for Animating Quadruped's Locomotion In 3D".He is specifically conducting research on generating procedural simulations of quadruped locomotion's. Mr. Bhatti is working as Lecturer in Department of Information Technology at Sindh University Jamshoro

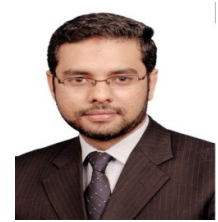

Pakistan. He has published many research papers in international journals and conferences.

### Dr. Asadullah Shah

Dr. Asadullah Shah is Professor at Department of computer science, Kulliyyah of information and communication technology, IIUM. Dr. Shah has a total of 26 years teaching and research experience. He has more than 100 research publications in International and national journals and conference proceedings. Additionally, he authored one book and currently editing another book. Dr. Shah has done his undergraduate degree in Electronics, Master's degree in Computer Technology from the University of Sindh, and PhD in Multimedia Communication, from the University Of Surrey, England, UK. His areas of interest are multimedia compression techniques, research methodologies, speech packetization and statistical multiplexing. He has been teaching courses in the fields of electronics, computers, telecommunications and management sciences.

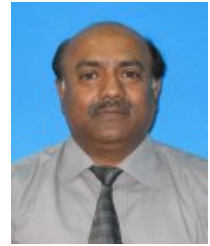

### Ahmad Waqas

Mr. Ahmad Waqas is PhD Scholar at Department of Computer Science, Faculty of Information and Communication Technology, International Islamic University Malaysia. He is working as Lecturer in the Department of Computer Science, Sukkur Institute of Business Administration Pakistan. He has been involved in teaching and research at graduate and post graduate level in the field of computer science for the last eight years. He has obtained his MCS (Masters in Computer Science) from University of Karachi in 2008 with second position in faculty. He did his MS (Computer Communication and Networks) from Sukkur IBA Pakistan. His area of interest is Distributed Computing, Cloud Computing Security and Auditing, Computing architectures, theoretical computer science, Data structure and algorithms. He has published more than 10 research papers in international journals and conference proceedings (IEEE and Scopus). He is working as technical committee member for different international journals and conferences.

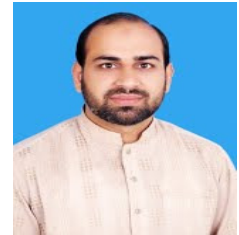

### Dr. Nadeem Mahmood

Dr. Nadeem Mahmood is working as Post-Doctoral Research fellow at Faculty of Information and Communication Technology, International Islamic University Malaysia. He is assistant professor in the Department of Computer Science, University of Karachi. He has been involved in teaching and research at graduate and post graduate level in the field of computer science for the last seventeen years. He has obtained his MCS and Ph.D. in computer science from University of Karachi in 1996 and 2010 respectively. His area of interest is temporal and fuzzy database systems, spatial database systems, artificial intelligence, knowledge management and healthcare information systems. He has published more than 20 research papers in international journals and conference proceedings (IEEE and ACM). He is working as technical and program committees' member for different international journals and conferences.

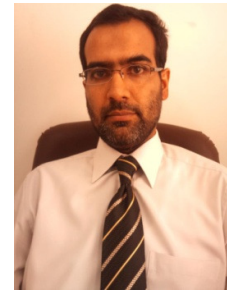